\documentclass[aps,pre,amsmath,amssymb,a4paper,twocolumn,
	showpacs,nofootinbib,notitlepage,10pt]{revtex4-1}

\usepackage{amsmath}    % need for subequations
\usepackage{graphicx}   % need for figures
\usepackage{verbatim}   % useful for program listings
\usepackage{color}      % use if color is used in text
\usepackage[hidelinks]{hyperref}   % use for hypertext links, including those to external documents and URLs
\usepackage{nicefrac}
\usepackage{lipsum}
\usepackage{upgreek}
\usepackage{notes2bib}
\usepackage{natbib}
\usepackage{xcolor}

\usepackage[normalem]{ulem}

\begin{document}

\title{Emerging Diversity in a Population of Evolving Intransitive Dice}
\author{Julius B. Kirkegaard}
\author{Kim Sneppen}

\affiliation{Niels Bohr Institute, University of Copenhagen, 2100 Copenhagen, Denmark}

\date{\today}

\begin{abstract}
Exploiting the mathematical curiosity of intransitive dice,
we present a simple theoretical model for co-evolution
that captures scales ranging from the genome of the individual
to the system-wide emergence of species diversity.
We study a set of evolving agents that interact competitively
in a closed system,
in which both the dynamics of mutations and competitive advantage
emerge directly from interpreting a genome as the sides of a die.
The model demonstrates sympatric speciation where new species evolve from existing ones
while in contact with the entire ecosystem.
Allowing free mutations both in the genomes and the mutation rates, 
we find, in contrast to hierarchical models of fitness,
the emergence of a metastable state of finite mutation rate and diversity.

\end{abstract}

\maketitle

\section{Introduction}
Evolution is the optimization scheme of the biological realm:
with the correct initial conditions
and ample time, random mutations and natural selection
are sufficient to ensure the emergence of highly complex organisms.
But the nature of what is being optimized is context-dependent.
Even for a fixed environment,
evolution does not necessarily have a single goal.
If we were to rerun the ``experiment'' of evolution,
it is virtually guaranteed this would
result in species distinct from those that are alive today.

Evolution of specific biological features
of a species depend on 
properties of the other species in its environment 
\cite{ehrlich1964butterflies,woolhouse2002biological}.
Such \textit{co-evolution} is
believed to be central to describe evolution on the large scale
\cite{KaufffmanBook, Bak1992, Sneppen1993,  Weitz2005, Xue2017}
as implied
by the \textit{Red Queen hypothesis} of Van Valen \cite{van1973new,liow2011red}.
Further, species interactions are 
not necessarily ranked, as observed among corals, plants and microbes \cite{jackson1975alleopathy, taylor1990complex,cameron2009parasite,kirkup2004antibiotic, Kerr2002}.
Many studies have been devoted to understanding and evolving such intransitive interactions,
ranging from molecular scale autocatalytic network 
\cite{eigen1978hypercycle, jain1998autocatalytic, segre2000composing}
to extensions of the competitive game of rock, paper and scissors
\cite{Kerr2002, Laird2006, Reichenbach2006, Reichenbach2007, Mathiesen2011, Mitarai2012, Dobrinevski2012, Ulrich2014, szolnoki2014cyclic, Levine2017, Soliveres2018}.

Non-hierarchical species dynamics can readily be studied for a set of species
whose interactions are fixed. 
Here, in contrast, we are interested in evolving systems
where intransitive interactions emerge \textit{ex nihilo}.
We suggest a minimal model for such a system consisting of
individuals that interact by rolling dice.

\begin{figure}[b]
    \centering
    \includegraphics[width=5cm]{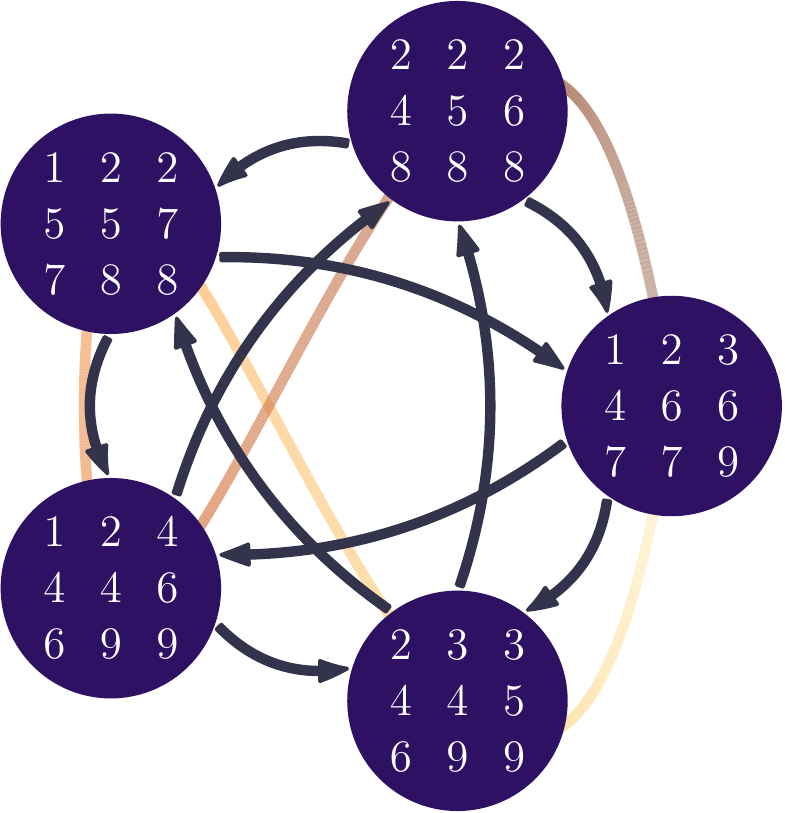}
    \caption{The intransitive interaction of five 9-sided proper dice.
    Direction of the arrows indicate domination.
    The graph formed from the interactions has two Hamiltonian paths,
  	one of which is indicated in the background.
  	In other words, out of the $4! = 24$ orderings of the dice, there are two
  	that form an intransitive loop of dominance.}
    \label{fig:graph}
\end{figure}

\vspace{1em}

\section{Model}
We define the characteristics of an individual in our system
by its genome as given by a list of $n$ integers.
We only consider competitive interactions between species
and settle these by interpreting
the genome of integers as the sides of dice that are rolled.
The outcome of a fight is stochastic, but certain dice will
tend to out-compete other dice.
For instance, a fight between $A = (3, 3, 3, 3, 3, 6)$
and $B = (2, 2, 2, 5, 5, 5)$ will typically be won
by $A$ despite $\sum_i A_i = \sum_i B_i$.
The probability of $A$ winning in the present example
is $n^{-2} \sum_i \sum_j [A_i > B_j] = 7/12$.
What makes this particular interaction interesting in the context
of competing species is the fact that we can introduce a species $C$ such that
both $B \succ C$ and  $C \succ A$, or succinctly:
$A \succ B \succ C \succ A$.
This is for instance the case for $C = (1, 4, 4, 4, 4, 4)$.
This intransitive behavior of dice is well-known \cite{Finkelstein2006, Conrey2016a},
but its applicability and simplicity for modeling co-evolution are unexplored.

A plethora of systems could be designed around the above interaction rule.
We choose to consider one of the simplest and study $k$
individuals that interact in a well-mixed scenario.
At each time step of our simulation, we let each individual randomly attack another.
Two individuals, $X$ and $Y$, are considered to belong to the same species if $\sum_i|X_i - Y_i| < \delta$,
in which case they will not fight.
Otherwise, the losing individual of the competition will be replaced by a copy of the winner.
On ties, a random winner is chosen.

\begin{figure}[tb]
    \centering
    \includegraphics{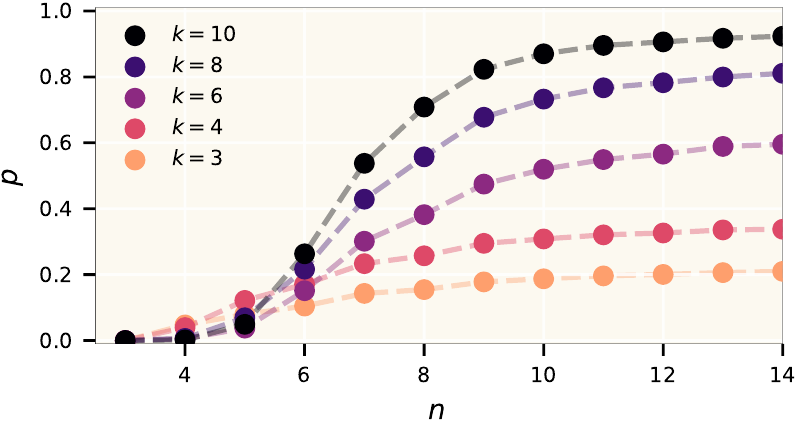}
    \caption{Probability of intransitive (Hamiltonian) $k$-loops in sets of $k$ proper dice
    as a function of the number of sides $n$ on the dice.
    Here, proper dice are sorted, has $\sum_i X_i = n(n+1)/2$ and $X_i \leq n$.
    A set of $k$ dice $\{ X_i \}$ has a $k$-loop if an ordering $\sigma$ exists
    such that $X_{\sigma_1} \prec X_{\sigma_2} \prec \cdots \prec X_{\sigma_k} \prec X_{\sigma_1}$.
    Each point is the result of averaging over $100,\!000$ Monte Carlo samples
    with uniform probability for each valid die.}
    \label{fig:count}
\end{figure}

\begin{figure*}[t]
    \centering
    \includegraphics{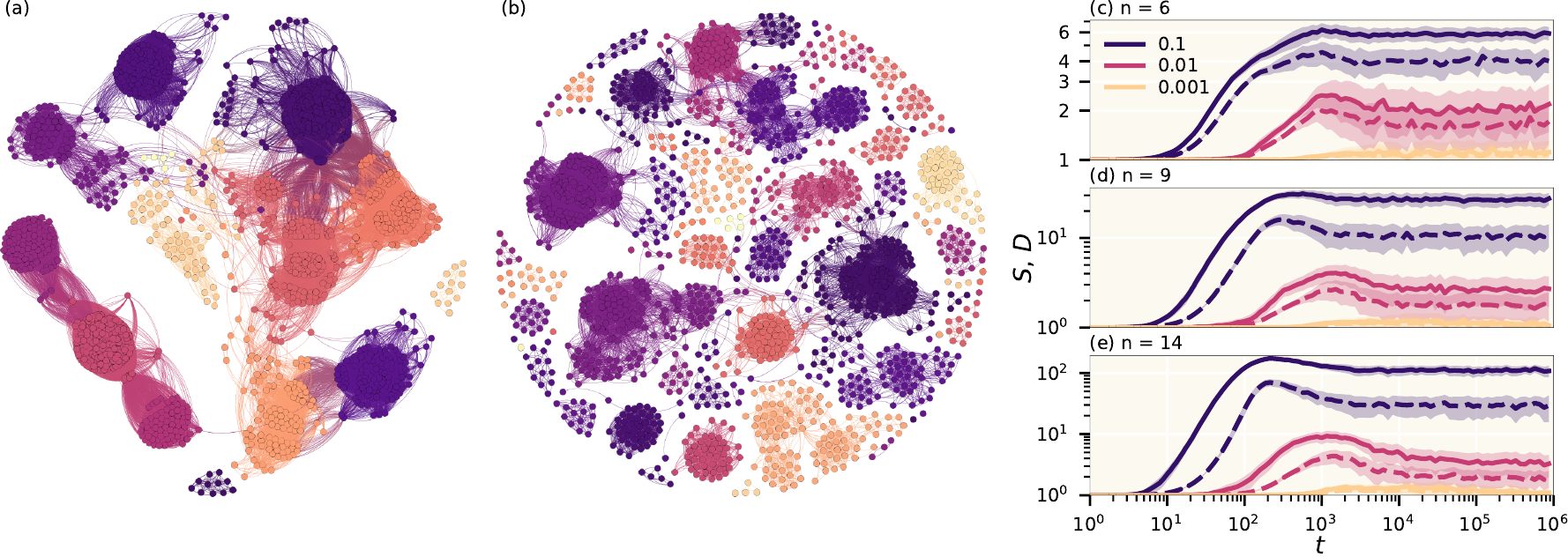}
    \caption{Speciation. (a--b) show $1,\!500$ individuals of the system with links between them
    if they consider one another the same species. (a) has $n = 9$ and (b) $n = 14$, and in both cases $\mu = 0.1$. Nodes are colored, for visualization purposes only, using the Leiden community detection algorithm \cite{Traag2019} on the undirected, unweighted graph.  (c--e) show species richness $S$ (solid lines) and diversity $D$ (dashed) stabilizing in a simulations of varying genomic complexity $n$ and mutation rate $\mu$ as indicated by the legend. Curves are averages over simulations with initial conditions of all dice equaling the standard die. All simulation were run with $\delta = 3$. $k=10,\!000$.}
    \label{fig:speciation}
\end{figure*}

A crucial novelty of our model is that the dice interpretation not only sets
the rules for interactions but also naturally provides a genome space in 
which mutations may occur. In our model, at each time step, each individual mutates with probability $\mu$.
A mutation event is the random change ($\pm 1$) of one of its genome digits.
As dice that are permutations of one another have identical competitive advantages,
we restrict our genome space to that of ordered dice.
Thus we disallow mutations that break the sorted nature of a genome.
For instance, $A$ may mutate to $(2, 3, 3, 3, 3, 6)$ but not to $(3, 2, 3, 3, 3, 6)$.
In effect this accelerates the dynamics of our system,
as most neutral competitions are avoided that would otherwise have to be settled
by stochastic extinction.
Further, naturally, it is universally better for a species to mutate up in sum rather than down.
We set a fitness ceiling by only allowing mutations that keep $\sum_i X_i \leq n \, (n + 1) / 2$
(the sum of the standard die) and any $X_i \leq n$;
the latter of which excludes a large of set of dice that are typically less competitive
and allows efficient enumeration of the set of allowed dice.

In total, we have thus created a mutation model where \textit{universal fitness} can be measured,
but where, occasionally, universally unfavorable mutations 
are preferred to adapt to coexistent competitors.

Instead of defining the interaction as a single roll of the dice,
one might consider competitions of $r$ rounds.
As $r \rightarrow \infty$, any slight competitive advantage will result in
certain overall wins. In this way, $r$ can be chosen to control the ruggedness
of the fitness landscape \cite{KaufffmanBook}.
Here, we limit our attention to $r = 1$.

\section{Probability of intransitivity}
Before delving into the dynamics of the model,
it is useful to have a feeling for the prevalence of intransitivity
in random dice.
Consider all dice such that $\sum X_i = n(n + 1) / 2$ and $1 \leq X_i \leq n$.
Fig. \ref{fig:count} shows the probability that $k$ such dice contains at least
one intransitive (Hamiltonian) $k$-loop as function of the number of sides of the dice $n$.
The plot shows the fact that if you choose $k = 3$ random dice (with $n < 15$ sides),
the probability that these dice interact intransitively is less than $\sim 20 \%$.
With a larger set of dice, not only do the probability of $k$-loops increase,
but so does probability of smaller sub-cycles
[see SI \cite{paperSI} for a version of Fig. \ref{fig:count} for loops of any size].
In Fig. \ref{fig:count} we only consider a small number of dice,
but very long intransitive cycles can also be found.
For instance, in the set of $k = 910$ allowed $9$-sided dice,
the longest possible cycle is at least $891$
(finding the precise length is NP-hard).

Thus, intransitivity is by no means rare, but (short) loops are not the norm either.
However, in the dynamics of our model, 
it is much more unlikely for a species to go extinct
in an intransitive loop than when species interact in a dominant manner.
Thus we expect one of two things to happen in the long run:
the system will be taken over by one species
or will be inhabited by a number of species that interact intransitively
and show oscillations.

There is also a large heterogeneity in the advantages of the different dice.
For instance, for $n = 6$, the die that has an advantage over most other dice
is $(1, 3, 3, 4, 5, 5)$, beating on average $\sim 60 \, \%$
of the other dice. The worst is $(1, 2, 2, 4, 6, 6)$, which is better than only $\sim 40 \, \%$.
In contrast, the standard die $(1, 2, 3, 4, 5, 6)$ has precisely a $50 \, \%$ chance of
beating any other proper die.
Similar conclusions can be made for all $n$.

We note that the precise statistics of Fig. \ref{fig:count} would be different if we
considered not only proper dice, as some dice
have many more permutations of their sides than others.
The qualitative conclusions drawn would remain similar, nonetheless.

\section{Finite mutation rate upholds species diversity}
Random, well-mixed ecosystems 
of many competing species interacting under demographic noise
are unstable \cite{may1972will}, and competitive exclusion
often leads to a collapse to only a single or a few surviving species 
\cite{hardin1960competitive, levin1970community, haerter2016food}.
This, naturally, also applies to the present dynamical system.
However, since the genome space of our model 
is not hierarchically organized,
a finite mutation rate can induce a
perpetual co-evolutionary arms race.

One complication in counting the number of species in a system
is due to the fact that the network of individuals is very unlikely
to organize into fully connected components.
This is a complication that is not unique to our system,
but indeed any problem related to speciation \cite{tucker2017guide}.
For each pair of individuals, our genome distance rule specifies if they 
belong to the same species.
Denote by $C_i$ the number of individuals that
individual $i$ is considered the same species as (including $i$ itself).
An effective measure for species richness is then given by $S = \sum_i C_i^{-1}$.
In the case of fully connected species, with no overlap between them,
this measure coincides with simply counting the number of species.
Likewise, we can define a species
diversity measure that also accounts for evenness as 
$D = k^2 \left( \sum_i C_i \right)^{-1}$.
This is equal to the number of species only if each species occupy the same
fraction of the entire system and thus small species contribute only
negligibly to its value.

Fig. \ref{fig:speciation} shows that both the mutation rate $\mu$
and the genome complexity $n$ (dice size) set the number of
species that a system of a certain size can maintain.
At low mutation rates, the system is dominated by 
a small cloud of individuals that form a quasispecies \cite{eigen1989molecular},
since, in this case, a single species can be locally
dominant and no individuals can escape this local optimum
at the low mutation rate.

At higher mutation rates, however, 
intransitive interactions appear and 
oscillatory dynamics of a 
high diversity system emerges.
The systems have intransitive loops of many lengths,
but the dynamics are dominated by short cycles ($\lesssim 5$ in the systems studied here).
This is demonstrated and 
studied in the SI by considering the mean-field Lotka--Volterra equations of the system.
In detail, the system behaves oscillatory with a frequency 
that remain relatively constant for multiple oscillation periods.
On long time scales, however, stochastic events
can change the dynamics, such as when a species in an intransitive loop
stochastically goes extinct or when an individual suddenly mutates to dominate
the existent intransitive interactions 
initiating a \textit{``punctuated equilibrium''}
event causing a sudden shift in oscillation frequency.

Fig. \ref{fig:speciation}(a--b) visualizes species connectivity
in the steady-state ecosystems that evolve from the dynamics of the model.
In these graphs, an edge is drawn between two dice if they consider
each other to belong to the same species.
Despite starting with a single species,
the system can evolve to one that has many species that are genomically disconnected.
While no single measure can capture the complexity of these inter-species connections,
running community detection algorithms on these graphs tend to find a number of clusters
in the same order of magnitude as our richness $S$ and diversity $D$ measures.
We note that for comparisons with other system sizes $k$,
the values of the mutation rates should be rescaled accordingly,
as the number of attempted mutation events per step is $\sim \mu k$.
Thus systems with a higher number of individuals can
support high diversity in spite of having a low
mutation rate per individual.

\section{Heterogeneous mutation rates}
Since a finite mutation rate is needed to maintain a finite diversity,
we have an ecosystem collapse if the mutation rate is taken to zero.
For instance, in a system of purely hierarchically interacting species,
the dominant species will prefer a low mutation rate thus leading to 
a collapse of ecosystem diversity.
In the present system, however, there is no global optimum, and 
a high mutations rate means quick adaptability and increases
the chance of an individual to out-compete others by an evolutionary advantage.
A high mutation rate is not strictly an advantage though, since it also means a
high rate of bad mutation events towards either locally or globally worse genomes.

\begin{figure}[tb]
    \centering
    \includegraphics{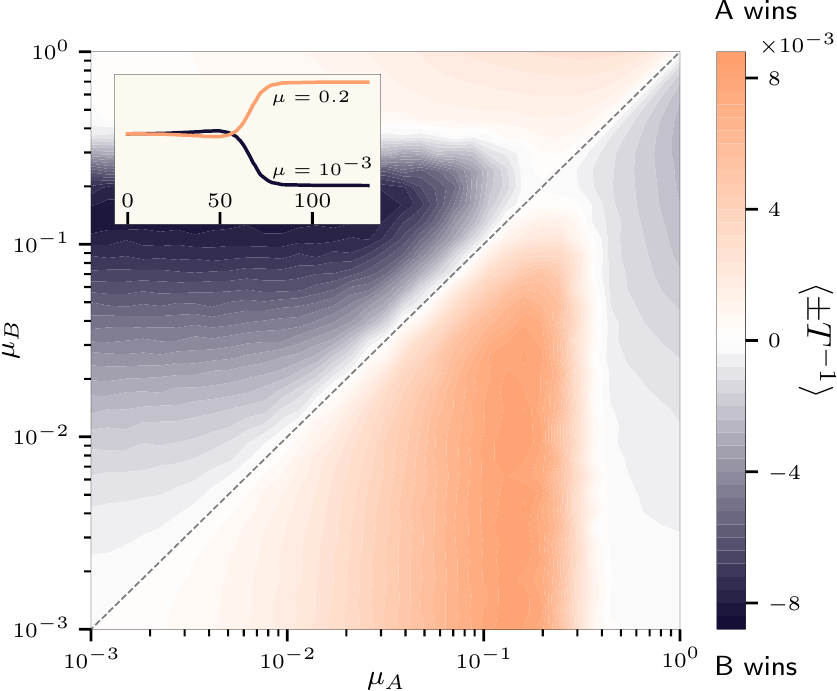}
    \caption{Annihilation statistics of two populations with different mutation rates.
    Inset shows a simulation of $2 \times 10,\!000$ dice with one half having $\mu_A = 0.2$ and 
    the other $\mu_B = 0.001$. The time of annihilation of population $B$
    is $T = 130$ time steps.  
    Main plot shows the average of $1/T$ measured with
    a negative sign if $B$ wins. High values thus indicate that $B$ tends
    to be annihilated and negative values that $A$ tends to be annihilated.
    Values near zero indicate a system dominated by stochastic extinction
    or one where annihilation takes a very long time. All dice were initialized 
    to $(1,1,1,6,6,6)$ at the beginning of the simulations.
    }
    \label{fig:1v1}
\end{figure}

In Fig. \ref{fig:1v1}(inset), we show the competition between two populations
with $\mu_A = 0.2$ and $\mu_B = 0.001$, respectively.
For early times, a low mutation rate gives an advantage because there is a low
rate of genomic decay and thus we see population $B$ winning initially.
This reflects the advantage of localizing a population
around a local fitness maximum over more diffuse quasispecies at 
higher mutation rates \cite{eigen1989molecular}.
However, at some point population $A$ finds a competitive advantage
over the slowly adapting population $B$ and annihilates the latter completely.
Varying $\mu_A$ and $\mu_B$, the average outcome of these scenarios is
shown in the main part of Fig. \ref{fig:1v1}.
The exact results depend on the initialization of the dice,
but in this case, we see that $\mu \approx 0.1$ is generally advantageous.

\begin{figure}[tb]
    \centering
    \includegraphics{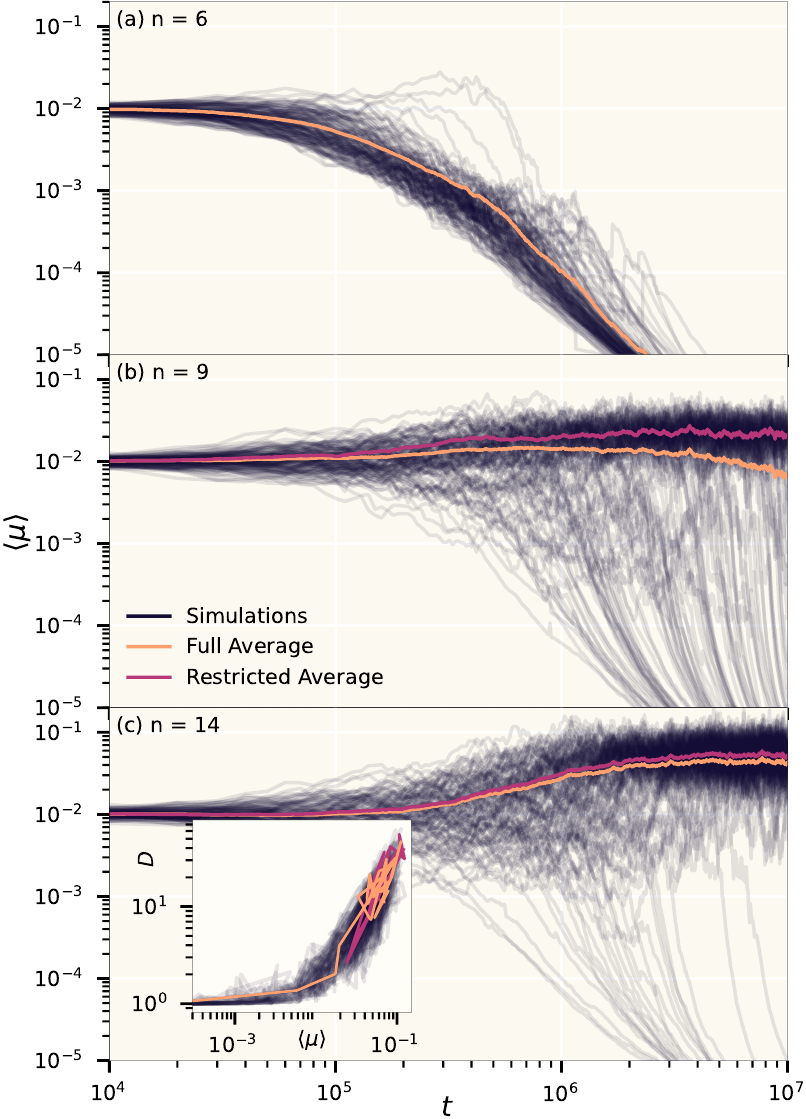}
    \caption{Mutating the mutation rate. Simulations for which each individual mutates
    its mutation rate with a 1 \% chance of a 1 \% change per time step (Gaussian multiplicative noise).
    (abc) Simulations for $n = 6, \, 9, \, 14$ showing the system average mutation rate
    $\langle \mu \rangle$ as a function of time steps. Orange lines show average over all simulations,
    whereas pink line shows an average over only those simulations that at $t = 10^7$ have $\langle \mu \rangle > 10^{-3}$.
    The full distribution of $\mu$ is narrowly peaked around $\langle \mu \rangle$ (see SI \cite{paperSI}).
    Inset in (c) shows, for each realization, the average mutation rate versus the system diversity $D$ (dark curves) for $t > 5 \cdot 10^5$, with two trajectories emphasized, one of which collapses towards $\langle \mu \rangle = 0$.
    All simulations use $ \delta=3$, but obtain similar results for other $\delta$. The fraction of systems in the high mutation state depends greatly on initial mutation rate, even for the $n=6$ die (see SI \cite{paperSI}).}
    \label{fig:mutatemutationrate}
\end{figure}

We complete the design of our model by, finally,
also permitting mutations in the mutation rate itself.
In a hierarchical setting, this would lead to the immediate collapse 
of both the mutation rates and diversity.
Fig. \ref{fig:1v1} indicates, however, that in the present system
there is also an advantage to having a finite mutation rate. 
Each panel in Fig. \ref{fig:mutatemutationrate} 
show realizations of a system of individuals,
all initialized with the same genome and an initial mutation rate of $\mu = 0.01$.
We observe two distinct outcomes:
most trajectories reach a metastable state with a high mutations rate
(die size $n=6$ never reaches this state,
but would do so if we instead started with initial $\mu=0.1$, see SI \cite{paperSI}),
and some that decay towards zero mutation rates.

A characteristic of the metastable state is the rare but sudden decay events of
both mutation rate and diversity of the entire system [Fig. \ref{fig:mutatemutationrate}(b--c)].
As the mutation rate decreases, it becomes less and less likely
to escape the low diversity situation thereby creating a positive feedback 
loop for decreasing the mutation rate even further.
Once collapsed, very large perturbations are needed to bring the system
out of this situation.
Even changing the mutation rate of half of the collapsed system to $\mu = 0.1$
is typically not enough to return to the high diversity state [see SI \cite{paperSI}].
In contrast, for the surviving high diversity ecosystems
the average number of mutations that separate two random individuals
is large. For instance, $\langle \sum_i |X_i - Y_i| \rangle_{X, Y} \approx 12$ for $n = 14$,
which is about $2/3$ of the average obtained between random dice.

\section{Perspective}
We have presented a theoretical model
that, on one hand, is exceedingly simple to define,
and at the same time successfully describes a host
of complex phenomena related to co-evolution.
At fixed, finite mutation rates,
the model permits a state of finite diversity
in co-evolutionary balance.
For hierarchically interacting systems, allowing mutations in the mutation rates themselves,
will lead to an ecosystem collapse.
In contrast, we find a metastable state of finite diversity,
whose stability increases quickly 
with genomic complexity, measured by the number of sides of the dice $n$.

We only considered a well-mixed system, meaning that at all times
each individual could meet any other individual.
Introducing space to the model, e.g. putting the agents on a lattice, 
will most likely stabilize the observed effects even further;
with spatial dynamics, intransitive relations will decay very slowly \cite{Mitarai2012},
and thus the rate of species extinctions decreases.
Furthermore, speciation events should
increase in frequency as space allows for 
transient allopatric speciation.

Precise quantification of intransitivity in the dynamical system
is another interesting avenue for further research:
despite being dominated by few intransitive cycles,
a static view of the systems at any given time will not reveal the dominance
of these cycles as at least one species will have a low population
count due to the oscillatory Lotka--Volterra-like dynamics imposed by the dynamics of
the intransitive loops.

In conclusion, from the simple rules of competing dice
emerge a natural balance of mutation rates and diversity.
Too high a mutation rate risks genomic decay and the disintegration 
of quasispecies: \textit{``mutate and die''}.
Too few mutations are disfavoured in analogy to
the Red Queen hypothesis: \textit{``mutate or die''}.

\vspace{1em}

\begin{acknowledgments}
This project has received funding from the Novo Nordisk Foundation Grant Agreement NNF20OC0062047 and from the European Research Council (ERC) under the European Union’s Horizon 2020 Research and Innovation Programme, Grant Agreement No. 740704.
\end{acknowledgments}

\end{document}